\documentclass[10pt,journal]{IEEEtranTCOM}
\usepackage[english]{babel}
\usepackage[usenames]{color}
\usepackage[cp1250]{inputenc}
\usepackage{amsfonts}
\usepackage{amsthm}
\usepackage{graphicx}
\usepackage{epsfig}
\usepackage{mathrsfs}
\usepackage{amsmath}
\usepackage{algorithm}
\usepackage{algorithmic}
\usepackage{hyperref}

\theoremstyle{plain}

\begin{document}
\newcommand{\bea}{\begin{eqnarray}}
\newcommand{\eea}{\end{eqnarray}}
\newcommand{\be}{\begin{equation}}
\newcommand{\ee}{\end{equation}}
\newcommand{\beas}{\begin{eqnarray*}}
\newcommand{\eeas}{\end{eqnarray*}}
\newcommand{\bs}{\backslash}
\newcommand{\bc}{\begin{center}}
\newcommand{\ec}{\end{center}}
\newcommand{\sech}{\textrm{sech}}
\def\SC {\mathscr{C}}

\title{Time crystal $\phi^4$ kinks by curvature coupling\\ as toy model
 for mechanism of oscillations propelled by mass, like observed for  electron and neutrinos}
\author{\IEEEauthorblockN{Jarek Duda}\\
\IEEEauthorblockA{Jagiellonian University, Cracow, Poland, \emph{dudajar@gmail.com}}}
\maketitle

\begin{abstract}
Dirac equation requires $E=mc^2$ energy of resting particle, leading to some $\exp(-iEt/\hbar)$ its evolution - periodic process of $\omega=mc^2/\hbar$ frequency, literally propelled by mass of particle, confirmed experimentally e.g. for quantum phase of electron as de Broglie clock/Zitterbewegung (and its angular momentum), or flavor oscillations of neutrinos for 3 masses. Entities having energetically preferred periodic process already in the lowest energy state are recently searched for as time crystals.

To understand such mechanism of clock propulsion by mass itself, it would be valuable to recreate something analogous in simple models like wobbling kinks. There is proposed such toy model as 1+1D $(\phi,\psi)$ Lorentz invariant two-component scalar field theory, extending popular $\phi^4$ model by second component $\psi$ corresponding to such periodically evolving degree of freedom, which is coupled through powers of curvature $R=\partial_0 \phi\, \partial_1 \psi-\partial_1 \phi \,\partial_0 \psi$, as suggested by earlier 3+1D model~\cite{my}. This way kink spatial structure $\partial_x \phi\neq 0$ brings energetic preference for nonzero $\partial_t \psi$ time derivative, by energy minimization leading to periodic process of  $0<\omega<\infty$ frequency, as required for time crystals.
\end{abstract}
\textbf{Keywords:} topological solitons, $\phi^4$-model, kink, time crystal, de Broglie clock, zitterbewegung, neutrino oscillations
\section{Introduction}
In Schr\"{o}dinger equation quantum phase evolves $\exp(-iEt/\hbar)$ for some energy $E$, which going to relativistic e.g. Dirac equation requires to include rest mass $E=mc^2$, leading to some $\omega=mc^2/\hbar$ frequency oscillations literally caused by the mass itself. For electron, beside having fixed angular momentum which in field theory should be some rotation of the field (not point), such intrinsic periodic process is called de Broglie clock/zitterbewegung, and was observed experimentally~\cite{clock} as increased absorption of electrons by silicon crystal, through some resonance of electron clock with spatial lattice. For neutrinos analogously 3 masses were introduced~\cite{neutrino} to explain their flavor oscillations.

\begin{figure}[t!]
    \centering
        \includegraphics[width=9cm]{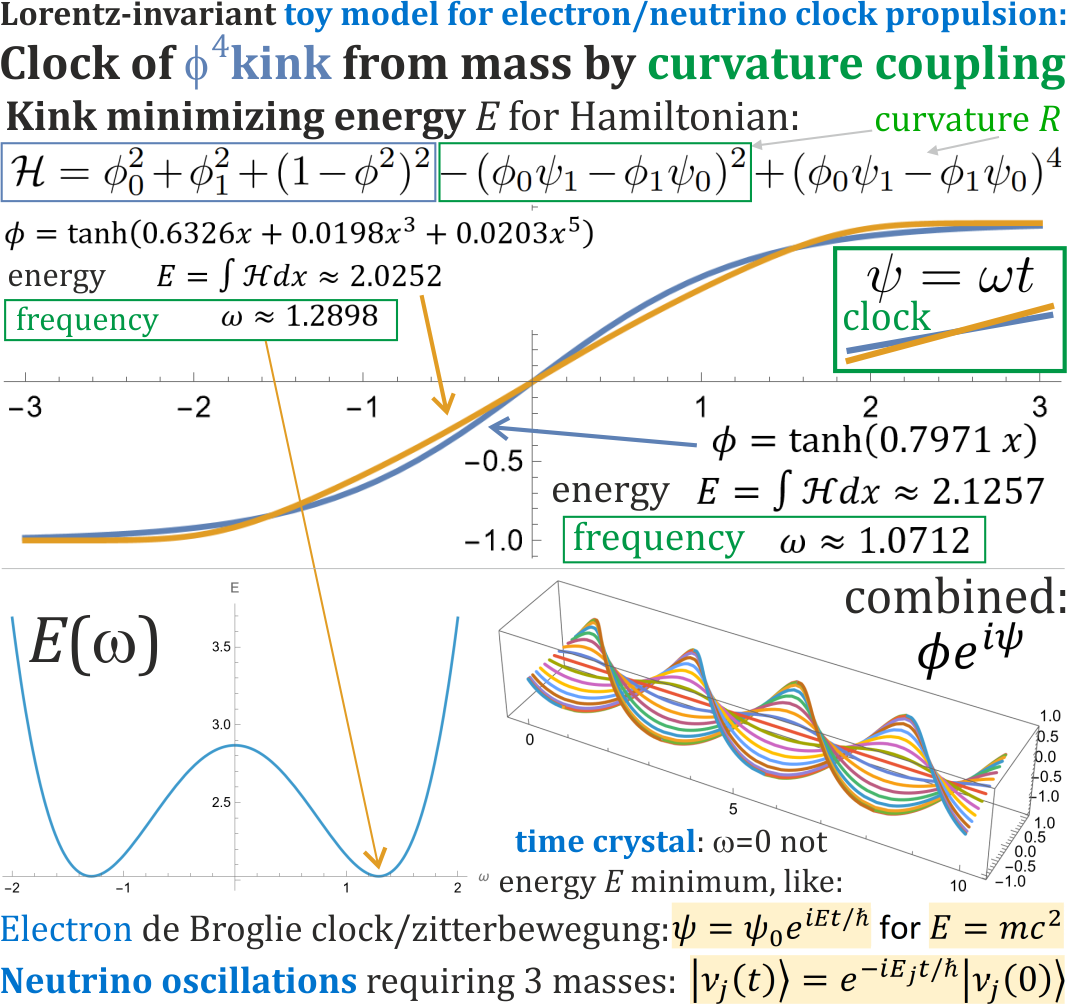}
        \caption{\textbf{Top}: numerically found deformation of  wobbling $\phi^4$ kink (code: \href{https://community.wolfram.com/groups/-/m/t/3398814}{community.wolfram.com/groups/-/m/t/3398814} ) for $\alpha=\beta=1$ case assuming $\phi=\tanh$(polynomial of $x$) and minimizing energy $E=\int_{-\infty}^{\infty}\mathcal{H}\,dx$ for coefficients of this polynomial, together with $\omega$ frequency. Coefficients for even powers of this polynomial have turned out zero, there are shown the odd ones up to $x^5$ as higher nearly did not bring energy reduction. \textbf{Bottom} left: energy frequency dependence for this optimized shape, with minimum for $0<\omega <\infty$, formally making it time crystal, like electron or neutrino. Right: visualized $\phi e^{i\psi}$ combining both scalar fields.
        }       \label{kink}
\end{figure}

It contradicts intuition that time derivatives should have positive energy contributions, hence energy minimization should slow oscillations down to zero. Somehow, at least for electron and neutrinos, their mass itself propels some intrinsic periodic process of e.g. quantum phase or flavors for neutrinos. In field theories mass is integral of Hamiltonian, suggesting it requires negative terms - preferring nonnegative time derivatives in presence of particles. While it seems there is no mainstream explanation of such propulsion mechanism, recently there are popular time crystals~(\cite{tc1,tc2}) which are supposed to have periodic process already in the lowest energy state, we will here realize for kink as in Fig. \ref{kink}.

\begin{figure*}[t!]
    \centering
        \includegraphics[width=19.cm]{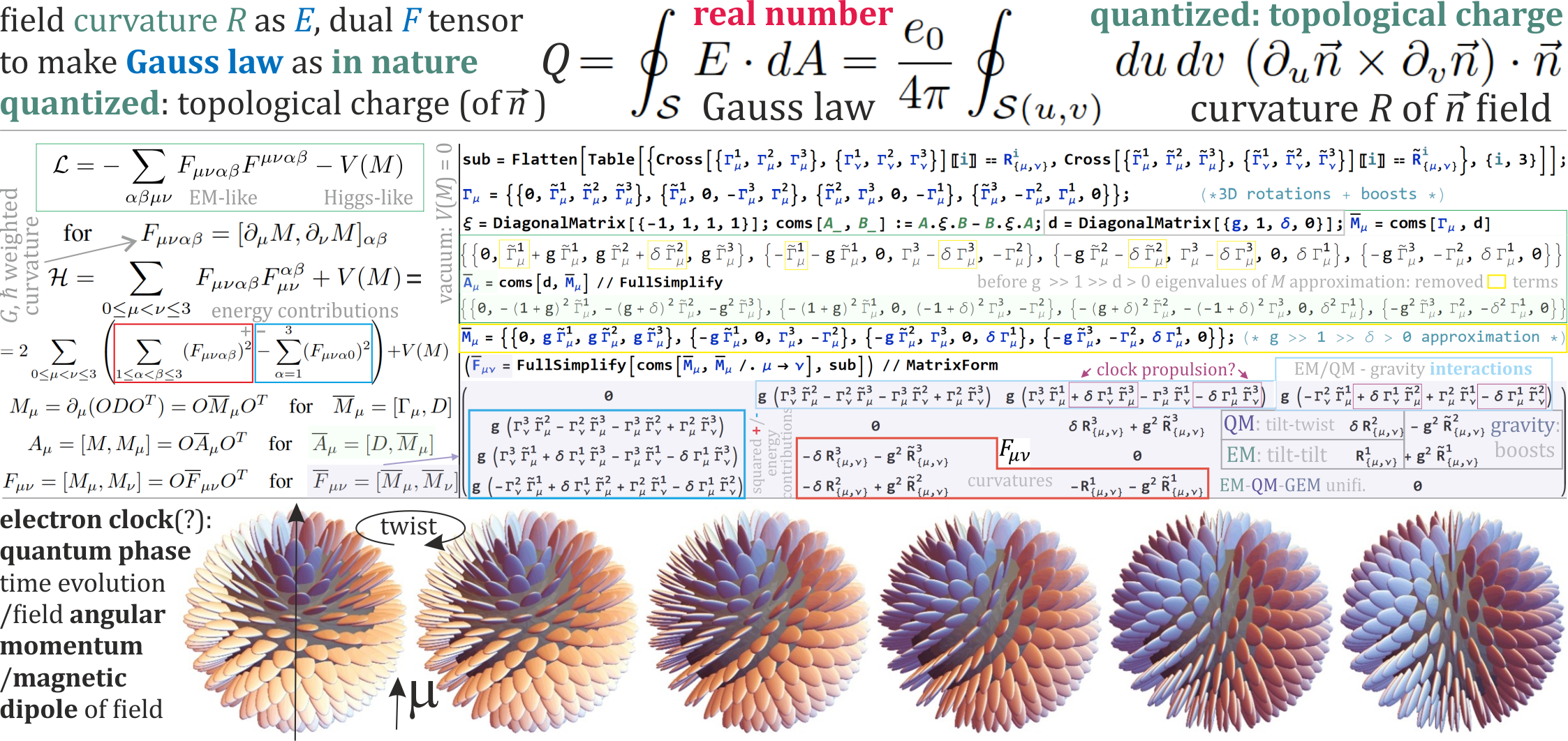}
        \caption{\textbf{Top}: the discussed curvature coupling  mechanism has automatically came from 3+1D model combining Skyrme and Landau-de Gennes approaches~\cite{my}, which starts with reparation of Gauss law: standard one returning charge being any real number, while in nature it is quantized. This disagreement can be repaired by defining electric field as curvature of some deeper field - this way Gauss law counts topological charge, which is quantized. It also regularizes central charge singularity to finite energy by Higgs potential, what deforms Coulomb force in very low distance/high energy in agreement with running coupling effect~\cite{faber1}. \\
        \textbf{Center}: its EM/Skyrme-like Lagrangian candidate using (dual) $F$ tensor as curvature $R$ of deeper field ($M$ tensor like in Landau-de Gennes model) to enforce charge quantization.
        While standard $F$ has 2 indexes, transforming to Hamiltonian with only positive terms ($B^2+E^2$), full curvature has 4 indexes, due to signature getting also negative squared curvature terms (marked blue), bringing the discussed mechanism propelling oscillations: $\Gamma_0$ in presence of particles. \\
        %The original motivation for curvature coupling from \cite{my}: 3+1D model combining Landau-de Gennes and Skyrmion approach, in which we interpret curvature $R$ of a deeper tensor field $M$ as dual $F$ tensor to recreate electromagnetism with built-in charge quantization as topological. In addition to EM from $S^2$ vacuum dynamics ($\Gamma^2,\Gamma^3$ generators), it also contains $U(1)\sim S^1$ twist ($\Gamma^1$) corresponding to quantum phase - due to $M$ eigenvalues having tiny $\hbar$-scale contribution to Lagrangian like in QED, and boost dynamics $(\tilde{\Gamma})$ leading to second set of Maxwell equations for \href{https://en.wikipedia.org/wiki/Gravitoelectromagnetism}{gravitoelectromagnetism} (GEM) approximation of the general relativity. Due to spacetime signature, EM-GEM interactions lead to subtle negative squared curvature Hamiltonian contributions (marked blue), providing energetic preference for both particle gravitational mass (local boosts $\tilde{\Gamma}$) and associated oscillations through nonzero time derivatives $(\Gamma_0)$. 
        \textbf{Bottom}: suggested electron field configuration: hedgehog of long axis for topological charge as quantized electric, recognizing twist corresponding to quantum phase - evolving as $\exp(-iEt/\hbar)$ for $E=mc^2$ required by Dirac, also to angular momentum: rotation of field (not point), and also magnetic dipole moment. }
       \label{Fmunu}
\end{figure*}

While perturbative QFT works on abstract point objects, it is only approximation of deeper nonperturbative view working on field configurations. Stable localized self-reinforcing field configurations are generally called solitons, topological if stabilized by topological constraints. They are one of nonperturbative approaches, hence have many more similarities with particle physics, like pair creation and annihilation, having charge quantization as topological - for which for example in liquid crystals there are observed long-range e.g. Coulomb-like interactions~\cite{coulomb}. Interpreting curvature of vector/tensor field as electric field like in Fig. \ref{Fmunu}, Gauss law counts its topological charge - allowing to recreate electromagnetism with built-in charge quantization as topological (Faber's approach~\cite{faber,faber1}). Extending to complete 4-index curvature~\cite{my} $F^*\sim R$ with $F_{\mu\nu\alpha\beta}F^{\mu\nu\alpha\beta}$ Lagrangian term, due to spacetime signature Hamiltonian obtains also subtle negative squared curvature terms, exactly as needed to prefer periodic process by energy minimization for topological solitons due to their mass itself.

While the above curvature coupling comes from complicated 3+1D tensor field combining Landau-de Gennes with Skyrmion model, to gain intuitions this article extracts simplified clock propulsion mechanism into much simpler 1+1D Lorentz invariant scalar $(\phi,\psi)$ two-component model, inspired also by recently popular this type wobbling kinks models~(\cite{two1,two2}).

\section{$\phi^4$-model curvature coupled with phase $\psi$}
Consider popular real scalar $\phi^4$ model~\cite{kinks} in 1+1D with $\eta_{\mu\nu}=\eta^{\mu\nu}=\textrm{diag}(1,-1)$ signature: for potential $V=(1-\phi^2)^2$ of minima in $\{-1,1\}$ (or sine-Gordon with sinusoid potential). We would like to couple $\phi$ with a second real scalar field $\psi$ representing this (quantum) phase ($\exp(i2\pi \psi)$  winded modulo 1), for which we would like to get energetic tendency for time evolution (nonzero $\psi_0 \equiv \partial_0 \psi$) from presence of particle-like kink (nonzero $\phi_1\equiv \partial_1 \phi$). 

Finding such oscillation-propelling coupling is nontrivial. As mentioned, we will show that squared curvature $R^2$ coupling as in previous 3+1D model in Fig. \ref{Fmunu}, with negative energy contribution $\alpha$ provides such propulsion mechanism. However, it alone would lead to frequency $\omega \to \infty$ by energy minimization, what in complete 3+1D model would be prevented by positive spatial curvature terms and other particles. Preventing it in the discussed simplified 1+1D toy model is more difficult, for simplicity let us now do it by additional positive $\beta R^4$ coupling, finally assuming Lagrangian:
\be \mathcal{L}=\partial_\mu \phi\, \partial^\mu \phi - (1-\phi^2)^2- \alpha R^2 +\frac{\beta}{3} R^4\ee
$$\textrm{for curvature coupling:}\qquad R=\partial_0 \phi\,\partial_1 \psi-\partial_1 \phi\,\partial_0 \psi$$
\noindent $R=\phi_0\psi_1 -\phi_1\psi_0$ denoting derivatives with lower index.
\subsection{Lorentz invariance, Hamiltonian, Euler-Lagrange}
To verify $R$ is Lorentz invariant, let us apply Loretnz boost $\phi(\gamma(t-vx),\gamma(x-vt))$ for $\gamma=1/\sqrt{1-v^2}$. This way $\phi_t=\gamma(\phi_0-v\phi_1)$, $\phi_x =\gamma(-v\phi_0+\phi_1)$, the same for $\psi$, leading to $\phi_0\psi_1 -\phi_1\psi_0 = \phi_t\psi_x -\phi_x\psi_t$ invariance.\\

\begin{figure}[b!]
    \centering
        \includegraphics[width=9cm]{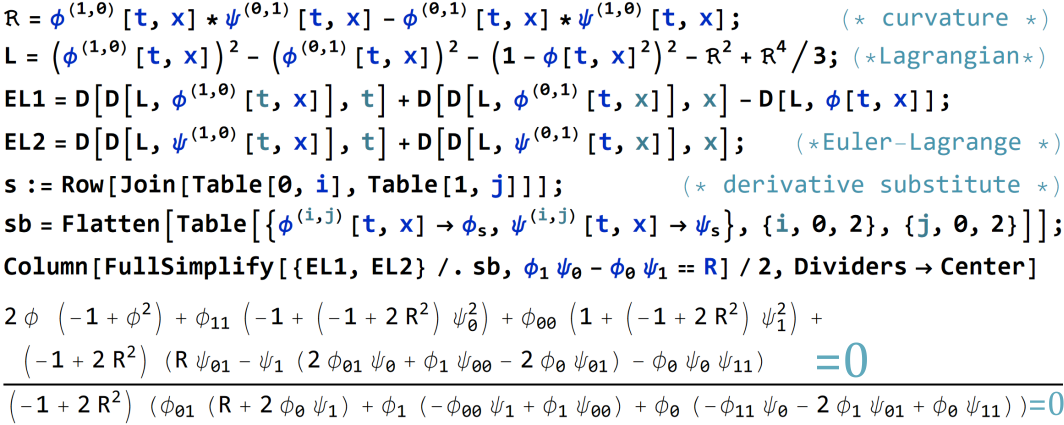}
        \caption{Mathematica search for Euler-Lagrange equations.
        }       \label{EL}
\end{figure}

Let us (Legendre) transform Lagrangian to Hamiltonian:
$$ \phi_0\frac{\partial R^p}{\partial \phi_0}+
\psi_0\frac{\partial R^p}{\partial \psi_0}-
R^p=(p-1) R^p $$
\be \mathcal{H} =\phi_0^2+\phi_1^2+(1-\phi^2)^2 - \alpha (\phi_0\psi_1-\phi_1\psi_0)^2+\beta(\phi_0\psi_1-\phi_1\psi_0)^4\ee
Euler-Lagrange equations are found by Mathematica in Fig. \ref{EL}.

\subsection{Energetically preferred frequency of kink's wobbling}
For $\phi=\textrm{const}$ there are no constraints for $\psi$ field, in practice it would be affected by other particles, also small $\phi$ perturbations would couple with $\psi$.

Let us consider kink-like solution for $\phi\equiv \phi(x)$ monotonously going from $\phi(-\infty)=-1$ to $\phi(-\infty)=1$, centered in zero: $\phi(0)=0$. This way $\phi_t=0$ hence there are no constraint for $\psi_1$, however, they would appear for perturbation. For simplicity let as focus on $\psi=\omega t$ linear phase evolution solution (modulo 1), leading to Hamiltonian:
\be \mathcal{H} = \phi_x^2 (1-\alpha \omega^2) + (1-\phi^2)^2 + \beta \omega^4 \phi_x^4 \label{ham}\ee
It allows to find frequency minimizing energy $E=\int_{-\infty}^{\infty}\mathcal{H}\, dx$:
\be \textrm{energetically preferred frequency:}\ \omega =\sqrt{\frac{\alpha}{2\beta}\frac{\int_{-\infty}^{\infty}\phi_x^2\, dx }{ \int_{-\infty}^{\infty}\phi_x^4\, dx }}
\ee

Let us find its approximate behavior assuming standard $\phi^4$-model $\phi(x)=\tanh(x/w)$ kink, here for width $w=1$ it would be solution for removed coupling: $\alpha=\beta=0$, should be perturbed for small $\alpha,\beta>0$. Substituting $\phi(x)=\tanh(x/w)$:
$$\mathcal{H}=\frac{w^2(1+w^2-\alpha \omega^2)\,\sech(x/w)^4+\beta \omega^4\, \sech(x/w)^8}{w^4}$$

Integrals can be found with hypergeometric functions:
$$\int_{-\infty}^{\infty}  \sech\left(\frac{x}{w}\right)^4 dx = \frac{4}{3} w\qquad \int_{-\infty}^{\infty}  \sech\left(\frac{x}{w}\right)^8 dx = \frac{32}{35}w$$
Finally searching for minimum of energy $E=\int_{-\infty}^{\infty} \mathcal{H}\, dx$, the zeroing of derivatives necessary condition gives:
\be \omega=\sqrt{\frac{70\alpha}{96\beta - 35\alpha^2}}\qquad \qquad w= \sqrt{\frac{96\beta}{96\beta - 35\alpha^2}}\quad \ee
suggesting $0<\omega<\infty$ energetically preferred oscillations at least for small $\alpha>0$ and $96\beta > 35 \alpha^2$. For example for $\alpha=\beta=1$ case, for which simple numerical optimization of energy for (\ref{ham}) Hamiltonian is shown in Fig. \ref{kink}.

\section{Conclusions and further work}
There was presented 1+1D toy model and its very basic analysis demonstrating curvature coupling mechanism, which brings energetic preference for $\psi_0$ intrinsic time evolution of kink due to its $\phi_1$ spatial structure, making it a time crystal (minimizing energy for $\omega\neq 0$ as in Fig. \ref{kink}), and a candidate for mechanism of mass propelling oscillations e.g. for electron and neutrino, like in Fig. \ref{Fmunu} automatically appearing from electric field as curvature to make Gauss law count topological charge.

The presented analysis can be improved e.g. by behavior including small perturbations with $\phi_t \neq 0$, and interactions of such wobbling  kinks. Finally, it should be extended to higher dimensions. In 2+1D such intrinsic oscillations should lead to coupled "pilot" waves, what should allow e.g. to recreate hydrodynamical Casimir effect~\cite{casimir} having waves originally caused by external shaker, which could be replaced by intrinsic particle oscillations. Then we could try to recreate "walking droplets" experiments with classical wave-particle duality objects - leading to interference~\cite{c1}, tunneling~\cite{c2} or orbit quantization~\cite{c3}. Finally, in 3+1D we should get proper models of, among others, electron clock/neutrino oscillations, e.g. with liquid crystal-like view from \cite{my}.

\bibliographystyle{IEEEtran}
\bibliography{cites}
\end{document}